\documentclass[useAMS]{mn2e}  
\usepackage{times}

\def\heao1{{\it HEAO-1\/}}

\newcommand{\ltsima}{$\; \buildrel < \over \sim \;$}
\newcommand{\simlt}{\lower.5ex\hbox{\ltsima}}
\newcommand{\gtsima}{$\; \buildrel > \over \sim \;$}
\newcommand{\simgt}{\lower.5ex\hbox{\gtsima}}

\def\lesssim{\mathrel{\hbox{\rlap{\hbox{\lower4pt\hbox{$\sim$}}}\hbox{$<$}}}}
\def\gtrsim{\mathrel{\hbox{\rlap{\hbox{\lower4pt\hbox{$\sim$}}}\hbox{$>$}}}}

\def\msun{M$_{\odot}$}

\usepackage{graphicx}
\usepackage{txfonts}
\usepackage{rotating}

\title[The dust-to-stellar mass ratio in galaxies]{The dust-to-stellar mass ratio as a valuable tool to probe the evolution of local and distant star forming galaxies} 

  \author[F. Calura, F. Pozzi, G. Cresci, P. Santini, C. Gruppioni, L. Pozzetti, R. Gilli, F. Matteucci, R. Maiolino] {F. Calura$^{1}$\thanks{E-mail:
      fcalura@oabo.inaf.it}, F. Pozzi$^{1,2}$, G. Cresci$^{3}$, P. Santini$^{4}$, C. Gruppioni$^{1}$, L. Pozzetti$^{1}$, R. Gilli$^{1}$, 
    \newauthor F. Matteucci$^{5}$, R. Maiolino$^{6}$\\ 
(1) INAF, Osservatorio Astronomico di Bologna, Via Ranzani 1, 40127 Bologna, Italy\\
(2) Dipartimento di Fisica e Astronomia, Universit\'a degli Studi di Bologna, Viale Berti Pichat 6/2, 40127 Bologna\\
(3) INAF-Osservatorio Astrofisico di Arcetri, Largo Enrico Fermi 5, I-50125 Firenze, Italy;\\
(4) INAF-Osservatorio Astronomico di Roma, via di Frascati 33, 00078 Monte Porzio Catone, Italy\\
(5) Dipartimento di Fisica - Sezione di Astronomia, Universit\`a di Trieste, Via G. B. Tiepolo 11, 34131 Trieste, Italy\\ 
(6) Cavendish Laboratory, University of Cambridge, 19 J. J. Thomson Avenue, Cambridge CB3 0HE, UK\\}
\pagerange{\pageref{firstpage}--\pageref{lastpage}}
\pubyear{---}

\begin{document}

\maketitle

\label{firstpage}
\begin{abstract}
The survival of dust grains in galaxies depends on various processes. 
Dust can be produced in stars, it can grow in the interstellar medium and 
be destroyed by astration and interstellar shocks. 
In this paper, we assemble a few data samples of local and distant star-forming galaxies to analyse various dust-related 
quantities in low and high redshift galaxies, to study how the relations linking the dust mass to the stellar mass and star formation rate 
evolve with redshift. 
We interpret the available data by means of chemical evolution models for discs and proto-spheroid (PSPH) starburst galaxies. 
In particular, we focus on 
the dust-to-stellar mass (DTS) ratio, as this quantity represents a true measure of how much dust per unit stellar mass survives 
the various destruction processes in galaxies and is observable. 
The theoretical models outline the strong dependence of 
this quantity on the underlying star formation history. 
Spiral galaxies are characterised by a nearly 
constant DTS as a function of the stellar mass and cosmic time, whereas PSPHs 
present an early steep increase of the DTS, which stops at a maximal value and decreases 
in the latest stages. 
In their late starburst phase, these models show a decrease of the DTS with their mass,
which allows us to explain the observed anti-correlation between the DTS and 
the stellar mass. 
The observed redshift evolution of the DTS ratio shows an increase from $z\sim0$ to $z\sim1$, 
followed by a roughly constant behaviour at $1\lesssim~z~\lesssim2.5$. Our models indicate  
a steep decrease of the global DTS at early times, which implies an expected decrease of the DTS at 
larger redshift. 
\end{abstract}

\begin{keywords}
dust, extinction -galaxies: evolution; star formation - infrared: galaxies.
\end{keywords}

\section{Introduction}

Interstellar dust represents the solid component of the interstellar medium (ISM), and 
it strongly affects several properties of galaxies. 
First of all, dust grains absorb and scatter the optical and ultra-violet (UV) radiation in the well-known 
dust-extinction process. Most of the light absorbed by dust at UV and optical wavelengths 
is then thermally re-emitted at much longer wavelengths, in the infrared (IR) band, where 
the dust-reprocessed radiation represents the major contributor. 

Second, the depletion of refractory elements into dust grains has a considerable effect on the chemical composition of 
galaxies. In our Galaxy, the depletion levels are found to vary largely as a function of the density and temperature 
of the ISM (Savage \& Sembach 1996; Jenkins 2009). This is true also in high-redshift damped Lyman alpha systems observed in 
Quasar and Gamma-Ray Burst absorption lines, which show strong evidence of dust depletion in heavy element abundances 
(e. g., Vladilo 2002; Calura et al. 2003; Calura et al. 2009a). 

Third, dust grain can enhance radiative cooling of the ISM and thus directly affect the star 
formation history of galaxies, 
as well as the stellar initial mass fuction (MF) by inhibiting the formation of 
massive stars (e.g. Bekki 2013; Omukai et al. 2005).

In galaxies, 
the survival of dust grains is regulated by a series of 
constructive and destructive processes which take place on rather distinct time scales. 
Dust grains are thought to form in the lowest-temperature and highest density regions 
of stellar envelopes and of the ISM. Favourable conditions for dust formation are met 
in the rapidly expanding shells of the ejecta arising from supernova (SN) explosions, as well 
as in outer envelopes of asymptotic giant branch (AGB) stars, on time scales ranging from 
a few Myr up to several Gyr (e.g. Dwek 1998; Tielens 1998; Edmunds 2001; Morgan \& Edmunds 2003; Zhukovska 2014; Michalowski 2015; Dell'Agli et al. 2015). 
Dust grains are also destroyed in SN shocks (Barlow 1978; McKee 1989; Jones et al. 1996; Schneider et al. 2004; Silvia et al. 2010), on times as 
fast at $\sim\,10^4$ yr after the SN explosion (Gall, Hjorth \& Andersen 2011). 
Dust grains can also grow in the coldest phases of the ISM (Draine \& Salpeter 1979; Draine 1990; Dwek et al. 2007; 
Pipino et al. 2011; Valiante et al. 2011; Asano et al. 2013; Schneider et al. 2016), typically on typical timescales a few tens of Myr (Michalowski 2015), i.e. of the order of the lifetime 
for local molecular clouds (Dwek 1998). 
Direct evidence for dust accretion comes from the observed increase of the depletion levels at large interstellar densities and low temperatures (e.g. Savage \& Sembach 1996) 
and from the observed infrared emission of cold molecular clouds (e.g. Steinacker et al. 2009), characterised by the absence of small-grain emission. 
These features can be accounted for by the coagulation of small grains on and into larger particles. Indirect evidence of dust accretion comes also from the estimation of the grain lifetimes, 
which would be very small if no process could allow them to recondense and grow (McKee 1989; Draine \& Salpeter 1979). 

In the last few years, extragalactic surveys in the infrared (IR) and sub-millimetre (submm) bands allowed us 
to perform an accurate estimate of the dust mass budget in local and distant galaxies. 
Recent studies revealed that submm galaxies (SMG) at redshift $z>1$ generally show a larger dust content than local spirals and 
even ultraluminous infrared galaxies (ULIRGs; e.g., Santini et al. 2010; Fisher et al. 2014), hosting the most vigorous 
dust-enshrouded starbursts in the local Universe. 
Although a fraction of them has been resolved into multiple
sources by ALMA observations (ALESS survey, Hodge et al. 2013), 
these surveys have revealed the difficulty to account for the large dust masses observed 
in high-redshift galaxies with standard assumptions regarding dust evolution
(Morgan \& Edmunds 2003, Dwek, Galliano \& Jones 2007;  
Santini et al. 2010; Valiante et al. 2011; Rowlands et al. 2014), i.e. with choices for the most basic parameters 
regulating dust growth and destruction tuned by reproducing the content of dust 
in local galaxies, also including a standard (i.e. non top-heavy) stellar initial mass function.  
 
Even studies of quasar (QSO) hosts at higher redshifts ($z>6$) indicate the presence of large masses of
dust in the early Universe, suggesting that the dust buildup has occurred on timescales considerably lower than 
$\sim$1 Gyr (e.g. Maiolino et al. 2004; Wang et al. 2013; Calura et al. 2014), i.e. the age of the Universe at those redshifts. 
This implies that in these systems, the formation of dust was possible through processes likely faster than the lifetimes of intermediate-mass stars, 
characterised by initial mass $2\le m/$\msun$<8$ and which at these epochs are still on the main sequence, 
hence they did not have enough time to produce considerable amounts of dust grains (but see Valiante et al. 2009).  
Moreover, the large star formation rate values of QSO hosts imply large destruction rates (Mattsson 2011; Gall et al. 2011a; Calura et al. 2014), as it has been 
widely reckoned that supernovae (SNe), the most rapid stellar dust factories (Dunne et al. 2003), are more efficient dust destroyers than producers 
(Dwek 1998; Calura et al. 2008), and this outlines the counterbalancing, fundamental role of the interstellar growth 
to support such large dust masses (Mattsson 2011, Valiante et al. 2011; Pipino et al. 2011). 
On the other hand, the absence of dust in metal-free, low star formation objects either at $z>6$ and in the local universe 
(Fisher et al. 2014) imply a dichotomy between rapidly dust-rich and dust-free objects. 

A still open question is the possibility that the 
buildup of dust may occurr in lockstep with that of metals, as suggested by a recent discovery by 
Zafar \& Watson (2013) of a constant, nearly solar dust-to-metal ratio in gamma-ray burst and QSO absorbers, 
with very little dependence on column density, galaxy type or age, redshift, or metallicity. 
On the other hand, De Cia et al. (2013) find significant variations of the dust-to-metals ratio in 
both QSO and GRB absorbers, with the discrepancy between their results and those of Zafar \& Watson (2013)  likely due 
to an overestimate of the intrinsic dust content estimated from the line-of-sight $A_V$, as done by the latter authors. 
These significant variations of the dust-to-metal ratio vs metallicity might be explained by means of a non-universal 
ratio between the stellar dust and metal yields related to a dependance on metallicity of the dust yields (Mattsson et al. 2014a). 

In this paper, we assemble data sets from various IR and submm surveys across a wide redshift range ($0\le z\le 6$) and 
we use published results to gain a deeper insight on the evolution of the dust mass budget throughout cosmic history. 
To this aim, we consider various dust-related galactic scaling relation and assess how they evolve with redshift. 
Our study is mostly focused on the amount of dust which is produced in galaxies per unit stellar mass, 
on how this quantity depends on the stellar mass and 
what are the main physical processes that drive its cosmic evolution. 
The observational data sets are compared to chemical evolution models for galaxies of different morphological 
types, which include a detailed treatment of dust production (Calura et al. 2008) and which allow us to put 
strong constraints on the relation between the quantities studied in this work and the underlying galactic star formation history.

\begin{figure*}
\centering
\includegraphics[width=180mm,height=100mm,angle=0]{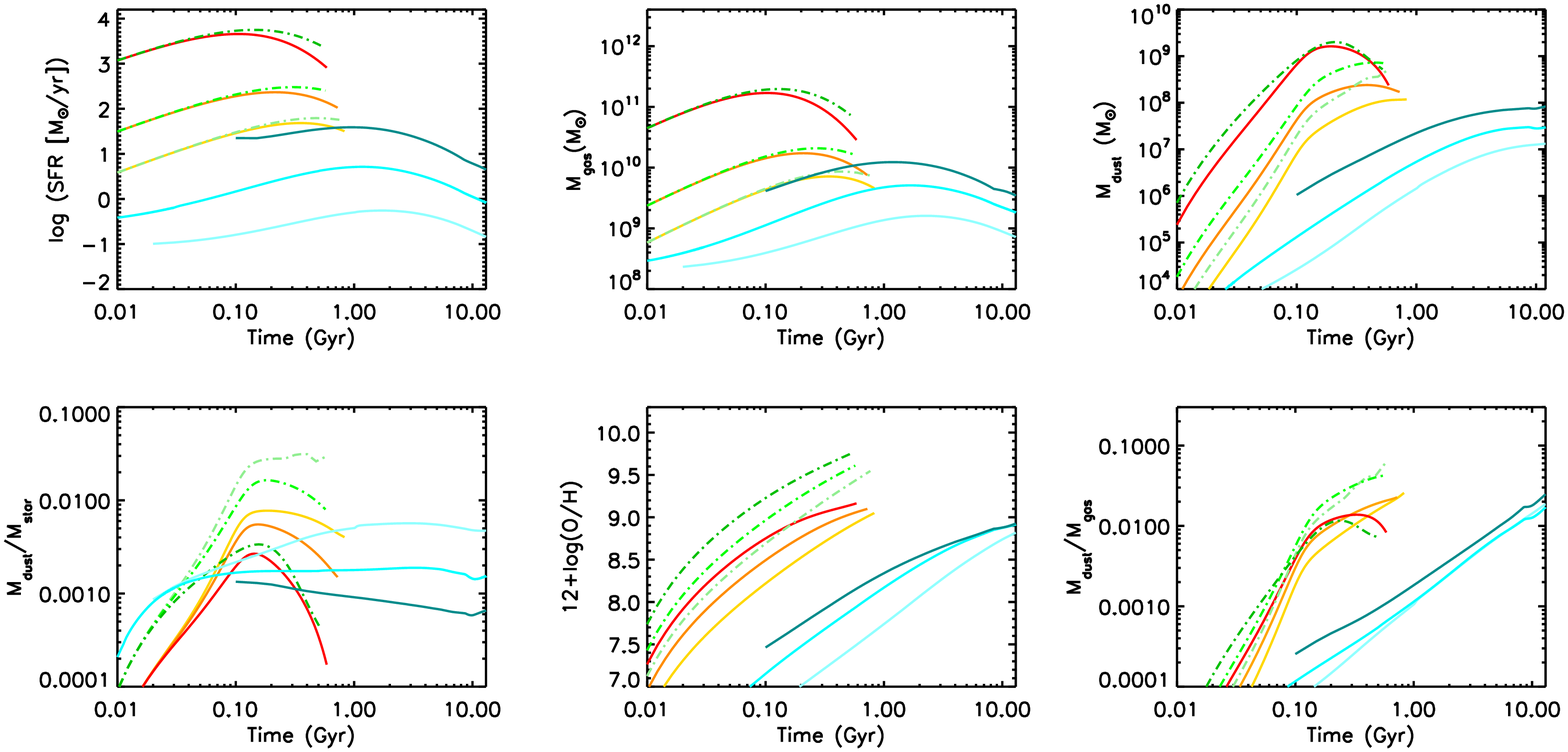}
\caption{Clockwise, from top left: star formation history, evolution of the gas mass, of the dust mass, 
of the dust-to-gas ratio, of 
the interstellar metallicity and of the
dust-to-stellar mass ratio as a function of time for our chemical evolution models described in Sect.~\ref{sec_mod}. 
In each panel, the yellow solid line,
the orange solid line and the red solid line represent the proto-spheroid models of 
baryonic mass $3 \times 10^{10}$, $10^{11}$ and $10^{12}$ M$_{\odot}$, respectively, 
computed with a standard, Salpeter (1955) IMF. 
The light green, green and dark green dot-dashed lines are for three PSPHs with the same baryonic masses as above but 
characterised by a top-heavy IMF (Larson 1998). 
The light cyan, cyan  and dark cyan solid lines represent models for a dwarf spiral, 
an intermediate-mass spiral and a M101-like spiral, respectively (Calura et al. 2009b).}
\label{fig0}
\end{figure*}

\begin{figure*}
\centering
\includegraphics[width=140mm,height=200mm,angle=0]{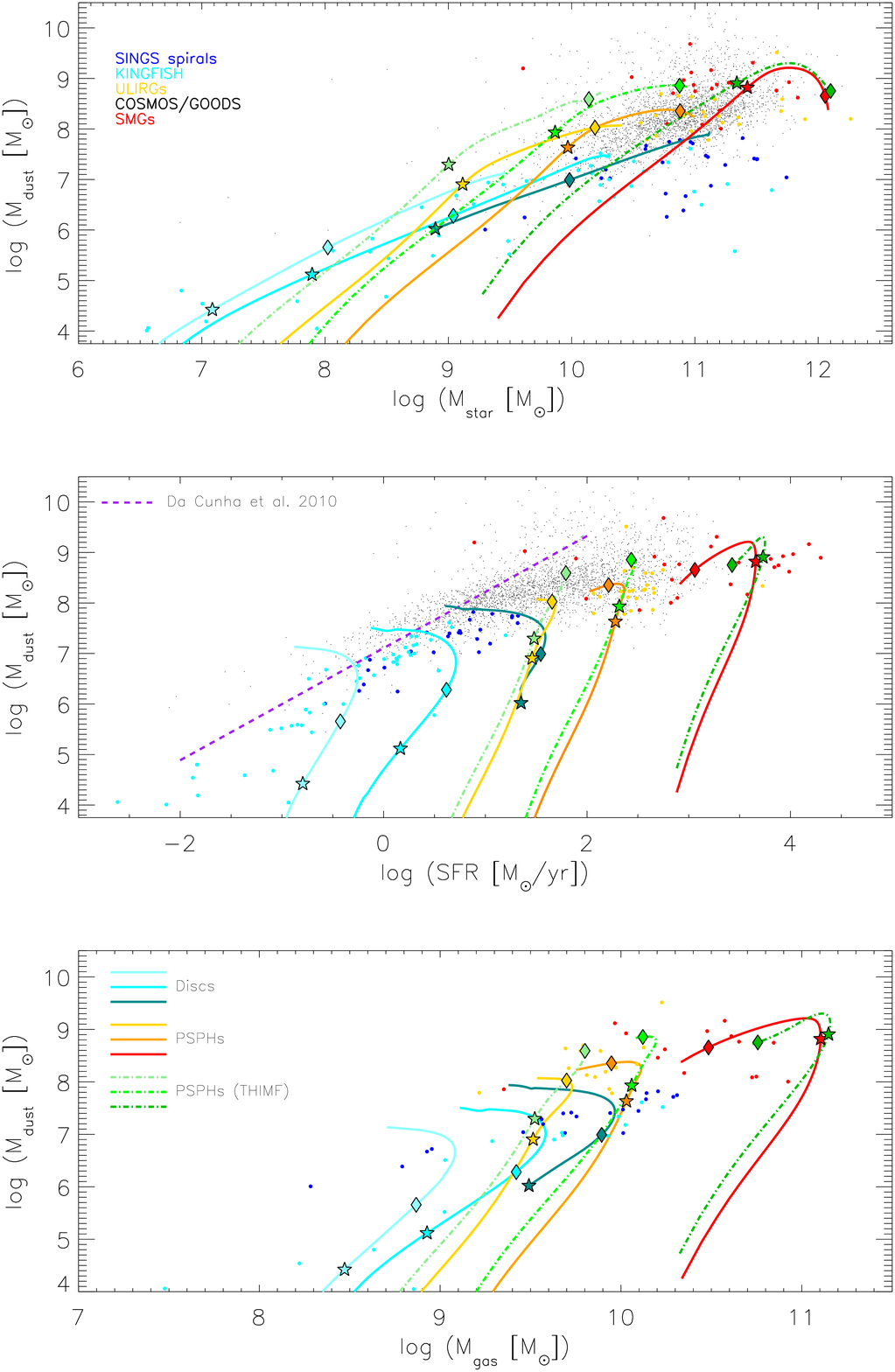}
\caption{Upper Panel: Dust mass as a function of the stellar mass for the observational data discussed in Sect.~\ref{sec_obs}. 
Symbols are colour-coded according to the different samples as follows. Blue circles are local 
spiral galaxies from the SINGS survey, light-blue circles are local 
galaxies from the KINGFISH survey, 
yellow circles are local ULIRGs 
and red circles are SMGs from Santini et al. (2010), whereas the black dots are galaxies from the COSMOS/GOODS survey. 
The curves represent the models described in Fig.~\ref{fig0}. 
The stars and diamonds plotted along each curve and with its same colour  
mark the evolutionary times of 0.1 and 0.5 Gyr, respectively. 
The central panel and the bottom panel show the observed and theoretical dust mass as a function of the star formation rate and 
the dust mass as a function of the gas mass, respectively. 
The purple dashed line in the central panel represents the 
the $M_{dust}$-SFR relation found in local galaxies by Da Cunha et al. (2010). 
All the other data and curves are as in the upper panel. 
}
\label{fig1}
\end{figure*}

\section{The observational data set}
\label{sec_obs}

In this paper, we focus on the stellar mass M$_{star}$, the dust mass M$_{dust}$, the gas mass
M$_{gas}$ and the star formation rate (SFR). 
The data set includes sources from the literature spanning
a wide redshift interval, for which some of the above physical
quantities are available. 
For a sub-set of data, novel estimates of some of
these quantities have been derived as explained below in Sect.~\ref{crossc}.

The data have been homogenized according to the following prescriptions: 
(i) a Salpeter IMF for the stellar masses and the SFR; (ii) a fit with a
modified black-body function for the dust masses (see Section 2.1). 
Finally, for the samples where both HI mass and molecular gas masses are available, 
the observed gas masses have been calculated as the sum of these two quantities. 
In the ULIRG and SMG samples, we assume that their total gas mass is coincident with their H2 mass.

\subsection{Dust Masses in the COSMOS and GOODS fields}
\label{crossc}
We derived novel estimates of the dust masses 
for galaxies in the COSMOS, GOODS-S and GOODS-N fields 
from the PEP Herschel survey (Lutz et al. 2011) and we used the
multi-wavelength dataset and source classification assembled by Gruppioni et
al. (2013) to derive the Herschel luminosity function. 

We have considered only the sources for which the observed flux $S_{\nu}$ was available 
for at least four photometric data points at
${\lambda}_{rest}>20$~${\mu}$m and all 
the sources dominated by an AGN were removed (i.e. classified in Gruppioni
et al. (2013) as type 1 and type 2 AGN).
A final sample of $\sim$2700 galaxies, in the redshift
range 0.01 $<z<$4.8, was obtained.\\

For these systems, as explained in Gruppioni et al. (2013), 
the stellar mass and the SFR were 
obtained through a detailed 
SED fitting in the optical-near IR bands, whereas  
the dust masses were estimated in this work by means of a modified black-body (MBB)
fit according to the prescriptions of Bianchi (2013). 
Under the assumption that all dust grains share a single temperature $T_d$ and that the dust distribution is optically thin, 
the dust mass M$_{dust}$ was calculated as 
\begin{equation}
M_{\rm dust} = \frac{D_{\rm L}^2 S_{\nu_{obs}} }{(1+z) \kappa_{\nu}B_{\nu}(T_{\rm d})}.
\end{equation}
In this equation, $B_{\nu}(T_{\rm d}$ is the Planck function,
 $D_{\rm L}$ is the luminosity distance of the object,
$S_{\nu_{obs}}$ is the flux observed at 250${\mu}$m (corresponding to a fequency of $1.2{~THz}$) and  
$\kappa_{\nu}$ and $B_{\nu}(T_{\rm d})$  were computed at $\nu=\nu_{obs}(1+z)$.
We have adopted $\kappa_{\nu} = 4.0 \left (\nu/1.2{~THz} \right)^{\beta}{\rm cm^2\,g^{-1}}$, 
with $\beta=2.0$ (see also Gilli et al. 2014). 
 
Bianchi et al. (2013) discussed how the bulk of the dust mass in galaxies 
can be robustly determined from the emission in the range $100~\mu$m$\le \lambda \le 500\mu$m, 
by fitting a single-temperature MBB to the data. However, 
the use of a MBB fit for the spectra to estimate the dust masses might 
seem questionable, as dust grains may present vastly different sizes and 
a large diversity of heating sources. 

Several authors have shown how stable fits to the data can be obtained also using a grain-temperature distribution (GTD; 
Kov\'acs et al. 2010; Magnelli et al. 2012), as it has been shown that dust grains have to be characterised by a GTD and 
its assumption deeply affects the estimate of the dust mass 
(see, e.g. Xie, Goldsmith \& Zhou 1991; Xie et al. 1993; Li, Goldsmith \& Xie 1999; Magnelli et al. 2012 and Mattsson et al. 2015). 
In general, the assumption of a GTD leads to dust mass estimates larger than those obtained by means of a MBB fit by factors of $1.5-2$ 
(e.g. Santini et al. 2014, Berta et al. 2016). For this reason, our estimates can be regarded as lower limits 
to the dust masses for the galaxies of our samples.

\subsection{Other data sets used in this work}

We have assembled two data sets at $z \lesssim 0.1$: 
a sample of spiral galaxies and a sample of ultra-luminous infrared galaxies (ULIRGs).

The spiral subsample includes 26 objects from the SINGS (Kennicutt et al. 2003), 
with full multi-wavelength photometry and also with submm data, useful for dust mass measurements. 
Estimates for the SFR, M$_{star}$ and M$_{gas}$ were
taken from Santini et al. (2010) and the HI and H2 mass estimates are 
from Kennicutt et al. (2003). 
The dust masses were re-calculated using a modified black-body fit 
\footnote{For these systems, the dust masses presented in Santini et al. (2010) were computed 
by means of the SED library generated by the GRASIL code (Silva et al. 1998).}.
The dust masses measured with a modified black-body fit result to be a
factor 2-3 lower than those estimated by the GRASIL code or
using the Draine \& Li (2007) templates (see Santini et
al. 2014). 

A subset of 46 spirals were selected from the KINGFISH sample (Kennicutt et al. 2011). 
For these sources, M$_{star}$ and M$_{dust}$ are from Skibba et al. (2011). 
The M$_{gas}$ estimates have been calculated from the sum of the HI mass (Walter et al. 2008) 
and the H2 mass (Leroy et al. 2009).

The ULIRGs sample consists of 24 systems, originally from Clements et al. (2010)
and, as for the spiral SINGs sample,
estimates for the SFR, M$_{star}$ and M$_{gas}$ 
were taken from Santini et al. (2010) by considering a CO-to-H2 
conversion factor ($\alpha$=4.3 $M_{\odot}$ (K km s$^{-1}$ pc$^{2}$)$^{-1}$) which applies to normal galaxies. 
No estimate for the HI mass is available for these systems. 
The dust masses have been re-computed by means of 
a fit with a modified black-body function, as explained earlier in Sec.~\ref{sec_obs}
 
At redshift $z>0$, we use the sample of 24 submillimeter galaxies (SMGs) presented by 
Santini et al. (2010). These sources are in the redshift range $0.5\le z\le 4$ 
and the estimates of M$_{star}$ and M$_{gas}$ were taken from 
Santini et al. (2010), whereas the dust masses were re-computed as described above for the SINGS and ULIRGs systems. 
We also consider one additional SMG source, SMMJ123600.15+621047.2 from  Greve et al. (2005).

Finally, for our study of the dust-to-stellar mass ratio as a function
of redshift (Sec. 4) we also consider the sample of QSO host-galaxies at $z>6$ 
from Calura et al. (2014), to which the reader is referred for further details. 
This sample shows that at $z\sim6$ a significant amount of dust is already present in several IR-luminous QSO hosts. 
A transition in dust production could have occurred at slighly larger redshifts, perhaps somewhere between $z\sim6$ and $z\sim7$,  
given the compelling absence of dust found in a couple of systems at $z>7$ (Ouchi et al. 2013; Inoue et al. 2016). 

For the data of Calura et al. (2014), the dust masses have been calculated from the FIR fluxes by means of Eq. 1, and the dynamical masses 
were estimated on the basis of either CO or [CII] line measurements (see, e.g., Wang et al. 2013). 
In the sample of Calura et al. (2014), only a few systems 
present molecular and/or atomic gas measurements (Walter et al. 2003,  Carilli et al. 2007, Maiolino et al. 2007, Wang et al. 2010; Wang et al. 2011a; Wang et al. 2011b;  
Venemans et al. 2012; Wang et al. 2013; Willott et al. 2013). 
Here we consider the systems for which it was possible to calculate 
the stellar mass as 
\begin{equation}
M_{star}=M_{dyn}-M_{gas}, 
\label{eq_mdyn}
\end{equation}
i.e. the systems characterised by $M_{dyn}>M_{gas}$. Here $M_{gas}=M_{HI}+M_{H2}$ and $M_{HI}$ and $M_{H2}$ are the neutral gas mass and the molecular gas mass values, as traced by 
the [CII]158$\mu$m emission line and the CO emission lines, respectively.
If one of $M_{HI}$ or $M_{H2}$ were not available, the unavailable quantity was estimated from the other one assuming 
a constant $M_{H2}/M_{HI}\sim 5$, equal to the average value computed from those systems for which both quantities 
were derived. 
To calculate the stellar masses for these systems, we neglect the presence of dark
matter. The contributions of supermassive black holes and dust were found to be minor compared to those of gas and
stars, thus neglected.

\section{Chemical evolution models}
\label{sec_mod}
The observational data sets described in Sect. \ref{sec_obs} are compared with results from chemical evolution models 
for proto-spheroidal (PSPH) galaxies and spiral, disc galaxies. 
In this section, we present a brief description of the chemical
evolution models used in this work. 
More detailed descriptions of the models can be found in 
Calura et al. (2009b), Schurer et al. (2009), Pipino et al. (2011) and Calura et al. (2014). \\
Elliptical galaxies are assumed to originate from 
the rapid collapse of a gas cloud with primordial chemical composition. 
During this 'proto-elliptical' phase, this rapid collapse triggers an intense and rapid star formation process, 
which lasts until the onset of a galactic wind, 
powered by the thermal energy injected by stellar winds and type II and type Ia SN explosions. 
At this time, the gas thermal energy equates the gas binding energy and all the residual 
interstellar medium is assumed to be lost. After this time, the galaxies evolve passively. 
The models considered in this work are designed to describe three PSPH galaxies of baryonic mass 
$3 \times 10^{10}$ $M_{\odot}$, $10^{11}$ $M_{\odot}$ and $10^{12}$ $M_{\odot}$. \\
We also consider a suite of chemical evolution models for spiral galaxies  
of three different baryonic masses, spanning from $\sim 2 \times 10^{9}$ $M_{\odot}$ to $10^{11}M_{\odot}$ (Calura et al. 2009b). 
In such models, the baryonic mass of any spiral galaxy is dominated by a thin disc of stars and gas in analogy with the Milky Way. 
The disc consists of several independent rings, 2 kpc wide, without exchange of matter between them. 
The  disc formation occurs 'inside-out', i.e. the timescale for disc formation increases with the galactocentric distance (Matteucci \& Francois 1989). 
In our models we assume that the 
efficiency of star formation is higher
in more massive objects that evolve faster than less massive
ones, a behaviour known as 'downsizing' (e.g., Cowie et al. 1996; Matteucci 1994; Recchi et al. 2009). 

All the models for spiral discs and for PSPHs include stellar 
dust production, mostly from core-collapse supernovae and intermediate-mass stars, restoring significant amounts of dust grains during the AGB phase. 
In this work, for all the refractory elements we consider the set of metallicity-independent dust condenstation efficiencies presented by 
Dwek (1998). For a discussion on the effects of metallicity on stellar dust production and its impact on galactic chemical evolution models, see 
Gioannini et al. (2016). \\
Dust destruction in SN shocks and interstellar dust accretion are also taken into account as described in Calura et al. (2008). 
The evolution of elliptical galaxies is followed during their star-forming phase only, i.e., in our scheme, 
before a galactic wind may occurr. However, it is worth stressing that also galactic outflows may represent an efficient mechanism to remove 
substantial amounts of dust from the interstellar medium (Calura et al. 2008; Feldmann 2015).\\

In Fig.~\ref{fig0} we show for our models, from the top-left panel  and in clockwise sense, 
the star formation histories along with the time evolution of the gas mass, of the dust
mass, of the dust-to-gas ratio, of the
metallicity (here traced by 12 + log (O/H)), and of the dust-to-stellar
mass ratio. 
The results for the PSPH models were computed both assuming a standard, Salpeter (1955) IMF and a Larson (1998), top-heavy IMF (THIMF).  
For the spiral models, we adopt isntead a Scalo (1986) IMF (see Sect.~\ref{sec_imf}).\\

The downsizing character of the PSPH models is
visible from the more extended SFHs in the less massive systems.
The SF is assumed to stop as soon as the conditions for the onset of
the galactic wind are met, i.e. when the thermal energy of the ISM
balances its binding energy.\\
In these models, the assumption of a THIMF generally produces larger dust masses (Gall et al. 2011a), larger metallicities (up by a factor of 
$\sim 0.5$ dex) and larger dust-to-stellar mass (DTS) values than those achieved with a Salpeter (1955) IMF. \\
The three spiral models are characterised by lower star formation rates and more prolonged
star formation histories than PSPHs. 
As the spiral models are multi-zone, each of the quantities plotted in Fig.~\ref{fig0} represent the mass-weighted averages
on the whole disc as described in Calura et al. (2009b) 
and considering only regions located at galactocentric distances $3$ kpc$<R<8$ kpc. 
As outlined by the recent 
Herschel Exploitation of Local Galaxy Andromeda (HELGA), local disc galaxies are known to exhibit dust-to-gas and dust-to-metals gradients (Smith et al. 2012; Mattsson et al. 2014b). 
Even if such gradients are likely to be present also in high-redshift galaxies, in this work we assume that 
for each object of our sample, the observational 
quantities we compare our results with are luminosity-weighted values representative of the galaxy as a whole. 
Such an assumption may be justified by considering that, 
as the innermost parts are the densest and the most luminous, the average quantities 
computed within the regions located at radii R$<8$ kpc could 
be regarded as representative of the entire discs (Calura et al. 2009b).

For the sake of consistency in the comparison with the observational gas masses described in Sect.~\ref{sec_obs}, 
the gas mass $M_{gas}$ refers to the total hydrogen mass present in the ISM in all models. 

The entire set of models used in this work was successful in reproducing a few basic properties  of 
galaxies and their evolution with cosmic time, such as the mass-metallicity relation (Calura et al. 2009b) and 
the dust abundance pattern in galaxies of different morphological types (Calura et al. 2008; Schurer et al. 2009; Pipino et al. 2011). Recently, the PSPH models have been matched with spectro-photometric models 
which include the emission from dust grains 
to successfully account for the evolution of the K-band 
and of the far-IR luminosity functions across the redshift range $0< z <3$ (Pozzi et al. 2015).

\subsection{Initial mass function}
\label{sec_imf}
In our standard PSPH galaxies, we adopt a Salpeter (1955) stellar initial mass function (IMF), 
i.e. a simple power law $\phi(m)\propto m^{-1.35}$ constant in space and time, 
with stellar mass lower limit $0.1~M_{\odot}$ and upper limit $100~M_{\odot}$. 
This choice assures that several observational constraints such as the average stellar abundances of present-day spheroids, 
the colour-magnitude diagram 
(Pipino \& Matteucci 2004) and their global metal content (Calura \& Matteucci 2004) are reproduced, as well as the 
metal content in clusters of galaxies and its evolution (see Renzini 2005; Calura et al. 2007). 

In spiral galaxies, the IMF is the one of Scalo (1986). This choice assures that the abundance pattern observed in local 
spiral and disc galaxies is well accounted for (e.g. Calura \& Matteucci 2004; Cescutti et al. 2007). \\
The use of this IMF is also motivated by various results showing that in general, 
the disc IMF must be significantly steeper than the cluster IMF, i.e. a universal Massey-Salpeter IMF with a Salpeter slope at the largest masses, 
i.e. at $m \ge 8~M_{\odot}$.
This result stems from the fact that the integrated IMF of the disc population can be obtained 
from a folding of the universal IMF with the star-cluster mass function 
(e.g. Weidner \& Kroupa 2005, Calura et al. 2010; Recchi \& Kroupa 2015), which is well described by a single power law (Lada \& Lada 2003).

We will also test the effect of a Larson (1998) IMF, of the form
\begin{equation}
\phi_{L}(m)\propto m^{-1.35} \exp(-m_{c}/m), 
\label{eq_lar}
\end{equation}
i.e. a power law modulated with an exponential function. 
The Larson (1998) IMF flattens below a characteristic stellar mass $m_{c}$ that may vary with time. 
By adopting a value of $m_c\sim 0.3$ M$_\odot$, of the order of the Jeans mass 
in local star-forming environments (Bate \& Bonnell 2005), the shape of the 
Larson (1998) IMF is qualitatively similar to the local, universal Kroupa (2001) or Chabrier (2003) IMF. \\
Recently, Calura et al. (2014) showed that a Larson (1998) 
IMF with $m_{c}=1.2 M_{\odot}$ in PSPHs produces an IMF richer in massive stars than the standard Salpeter IMF and allows one 
to account for the dust content of starbursts detected at $z\sim 6$. 
In high-redshift starbursts $m_{c}=1.2 M_{\odot}$ will be our reference value. 

As discussed in Mattsson (2011), in principle, 
since observations suggest that low and intermediate mass stars produce 
significantly more dust per stellar mass than high-mass stars, if stars were the only sources of dust a THIMF would result in a lower total stellar dust yield. 
However, also processes other than stellar dust production are at play (such as dust growth) and overall the adoption of a THIMF 
implies larger dust amounts than those achievable with a normal IMF.

\section{Results}
\label{Scal}
Various scaling relations for our
galaxy samples at different redshifts 
are plotted in Fig.~\ref{fig1}-\ref{fig3}, where 
the observational data sets described in Sect.~\ref{sec_obs}
are compared with theoretical results from chemical evolution models of galaxies of 
different morphological types as described in Sect.~\ref{sec_mod}. 
Combined together, the dust masses, the gas masses and the stellar masses  
allow one to have an insight on 
a few basic physical properties related to local and distant 
galaxies and on how some fundamental galaxy 
parameters evolve with redshift in the selected samples. 
The comparison with chemical evolution models 
is useful to interpret the data, since it can provide information 
on the possible morphology of the systems, and it also allows one to 
put constraints on their star formation history (Calura et al. 2009a). 

\subsection{Dust mass vs stellar mass}
\label{mdust_mstars}

As discussed in Santini et al. (2014), the positive correlation 
between  $M_{dust}$ and 
$M_{star}$ can be regarded as a consequence of the $M_{dust}$-SFR correlation 
combined with the $M_{star}$-SFR relation (i.e. the main sequence relation, see e.g., Elbaz et al. 2007)
However, to a first approximation, 
the correlation between $M_{dust}$ and 
$M_{star}$ can also be regarded as a consequence of the 
mass-metallicity (MZ) relation, if one considers that the refractory 
elements representing the constituents of the dust grains are 
mostly metals (e.g., Dwek 1998; Calura et al. 2008).
Clearly, 
in both the MZ and $M_{dust}$-$M_{star}$ diagrams, one has to consider that 
galaxies do not evolve as 'closed boxes', and that processes such as 
gas accretion and outflows 
can play important roles in determining both relations. 
Moreover, while the metal budget is regulated solely by the balance between stellar production 
and astration, the interstellar dust budget also depends on 
other complex mechanisms such as 
grain destruction in SN shocks as well as dust growth. \\
Even though the $M_{dust}$-$M_{star}$ relation alone does not allow us to 
determine the relative importance of all these processes, 
it clearly outlines that the growth of the dust mass has 
to occurr in lockstep with the increase of the stellar mass, and that this 
is true at any redshift. \\
As visible from the $M_{dust}$-$M_{star}$ panel of Fig.~\ref{fig1}, 
the trend characterising most of the observed local disc galaxies is consistent 
with the theoretical evolutionary tracks calculated for the spiral models, with the exception 
of a set of local SINGS and KINGFISH systems with large stellar mass ($log M_{star}/ M_{\odot} \gtrsim 10.5$) 
and low dust content ($log M_{dust}/ M_{\odot} \lesssim 7$). We checked the 
morphological classification of these high-mass, low-dust objects and we found that the majority ($> 60 \%$) 
of such systems belong to E, S0, SA0 and SB0 types, which are not represented by the suite of models used in the present 
work. 
Such low dust masses are instead in qualitative agreement with the values predicted 
for large-mass early-type galaxies with ages $>10$ Gyr, generally characterised by dust masses $>10^6 M_{\odot}$ (Calura 
et al. 2008). 

The observed stellar and dust masses of discs  
partially overlap also 
with the curves calculated for the lowest masses PSPHs, in particular 
in their earliest phases, i.e. when they are characterized by low
stellar masses and little dust content. 
As we will see  later in Sect.~\ref{mdust_sfr}, the key parameter required to better constrain the star formation history of these systems 
and to disentangle between different models is the SFR. 
During their early starburst phase, the PSPHs models are 
also characterised by a steeper slope of the  $M_{dust}$-$M_{star}$ relation than spirals. 
This feature clearly reflects the different star formation histories of the 
two model types, visible in Fig.~\ref{fig0}, where one can appreciate how the PSPHs show a considerably more rapid early 
increase of the dust masses with respect to spirals. 
This can be understood by considering that to a first approximation, during the earliest stages when both growth and destruction can be neglected, 
dust production scales as $dM_{dust}/dt \propto (y_{dust}- \frac{M_{dust}}{M_{gas}}) dM_{star}/dt$, where the 'effective dust yield' $y_{dust}$ accounts for the dust returned 
from a simple stellar population, which in principle is independent from the star formation history (although it may be a function of the metallicity as the dust condensation efficiency, see, e.g., 
Gioannini et al. 2016) whereas the dust-to-gas ratio $\frac{M_{dust}}{M_{gas}}$ is in turn strongly dependent on the star formation history (see Fig. ~\ref{fig0}). 

The local ULIRGs have both stellar and dust masses larger than local spirals, 
compatible with the theoretical tracks of the 
intermediate-mass PSPH model in the latest phases and of the most massive 
PSPH model during the earliest stages. 
The dust masses of SMGs are broadly consistent with 
the values predicted for the most massive elliptical model  
at masses $M_{star} > 10^{11} M_{\odot}$, i.e. at stellar masses larger than the observed ones. 

The COSMOS/GOODS sample includes a large variety of systems, most of them compatible with the PSPH models. 
This sample includes also several systems whose large 
dust content is difficult to account for 
by any galaxy evolution model. This fact outlines a problem 
already addressed in a recent study of the infrared and submillimetre 
properties of high-redshift galaxies (Rowlands et al. 2014), and confirms 
results already achieved in a few previous works, i.e. a dust mass produced per unit stellar mass 
significantly larger than what expected in standard models (Gall et al. 2011b; Calura et al. 2014, Michalowski 2015). 
Later in Sect.~\ref{disc}, we will come back on the main implications 
of this result. 

In the PSPH models, the use of a THIMF 
\footnote{For the conversion of the observed 
stellar masses from a Salpeter to a Larson (1998) top-heavy IMF with $m_{c} = 10 M_{\odot}$, 
Hijorth et al. (2014) suggest a rescaling downwards by a factor $\sim 3$, corresponding to -0.48 dex in logarithm (see also Dwek et al. 2011). 
In our case, we assume a considerably less extreme IMF characterised by $m_{c} = 1.2 M_{\odot}$, 
hence such a conversion should be regarded as an upper limit. In our case, a reasonable conversion is likely to be of 0.1-0.2 dex. Given the large dynamic range 
of stellar masses characterising the models and the data in this work, which spans from $\sim~10^6~M_{\odot}$ to $10^{12}~M_{\odot}$, in our analysis we ignore such a conversion.} 
alleviates the tension 
between the COSMOS/GOODS data presenting the most extreme dust mass values and the models. 
Only a minority of systems with $M_{dust} \simgt 10^{9} M_{\odot}$ 
is still unaccounted for by the models 
including a THIMF. 
The strong FIR emission of these objecs could be related to various effects, including the blending of multiple sources, lensing or variable 
syncrotron emission (Santini et al., in prep.). 
Additional data (especially with ALMA) are needed to securely rule out (or validate) 
such alternative scenarios.

\subsection{Dust mass vs SFR}
\label{mdust_sfr}

As outlined by previous studies, both local and distant galaxies show a tight and clear relation between 
dust mass and SFR (Da Cunha et al. 2010; Rowlands et al. 2012; 
Hjorth, Gall \& Michalowski 2014; Santini et al. 2014). 
The behaviour of such a relation is visible 
in the middle panel of Fig.~\ref{fig1}. 

The local spiral samples and the 
COSMOS/GOODS data set confirm the existence of a correlation between SFR and $M_{dust}$ 
although the slope seems different than the one found for local galaxies. 
It is worth stressing that the SMGs data set allows us to extend the $M_{dust}$-SFR  relation to SFR values $>10^3\,M_{\odot}/yr$, i.e. 
to values larger than those explored in previous studies (e. g., Hjorth et al. 2014). 

The SFR and $M_{dust}$ values observed in SMGs seem to confirm what already found in local galaxies and high-redshift galaxies, 
i.e. that the relation seems to bend over at very large SFR values (Hjorth et al. 2014). 
Moreover, in galaxies at redshift $z>0$, the $M_{dust}$-SFR  relation seems shallower 
than that found in local galaxies by Da Cunha et al. (2010). 
As seen in Fig.~\ref{fig1}, the largest dust masses observed in high-redshift systems such as, e.g., 
in SMGs, are of the same order of magnitude of the maximum dust masses of the 
local sample of Da Cunha et al. (2010), whereas the SFR values of such high-redshift systems are much larger.  \\
It is noteworthy that in the Da Cunha et al. (2010) sample, both dust masses and SFR values were 
computed in different ways than in the present paper. For instance, in that work 
the dust masses were computed with a multi-component model of the galactic spectra, 
and considering the emission from hot, warm and cold dust as well as 
a polycyclic aromatic hydrocarbons component. The different methods of calculation, as well as different 
spectral coverages in the samples (see Hjorth et al. 2014) could lead to 
systematic differences between the relation found by Da Cunha et al. (2010) and our results. 
A quantitative assessment of these systematic effects are beyond the aim of the present paper. 
The simultaneous study of the $M_{dust}$-$M_{star}$ and $M_{dust}$-SFR relations is very useful in order to 
disentangle among different possible star formation histories characterising our systems. 
In principle, this is similar to the combined study of the 
MZ and SFR vs metallicity relations performed by 
Calura et al. (2009b) to have fundamental hints 
on the nature of the galaxies building the MZ relation. \\
In the models, a clear distinction between discs and 
PSPHs is visible in the $M_{dust}$ vs SFR plot. 
Barring a small overlap between the lowest mass PSPH model 
and the largest mass spiral model, which concerns an extremely narrow SFR range, 
the two morphological types tend to occupy rather distinct regions of this diagram.  
The lowest star formation rates and dust masses characterize 
the evolutionary tracks of the spiral disc models, 
whereas the PSPHs occupy the highest-SFR, largest $M_{dust}$ regions. 

The KINGFISH and SINGS data are generally in good agreement with the spiral model tracks,  
which essentially removes the degeneracy discussed in Sect.~\ref{mdust_mstars}, which 
saw PSPHs and spirals presenting similar $M_{star}$ and $M_{dust}$ values during the earliest stages. 
Here, the lower SFR values observed in the KINGFISH and SINGS samples show a clear agreement 
with the spiral models and an inconsistency with any PSPH model. 

The majority of the ULIRGs data lie between the intermediate-mass and higher-mass PSPH models. 
The SMG data show a large spread in SFR. Four systems show SFR values lower than $10^2M_{\odot}/yr$, 
whereas the majority of them lie either between the intermediate-mass and largest-mass PSPH models, 
or rightwards of the largest mass PSPH model. For all the data sets, globally, in terms of SFRs, 
there is a good correspondence between 
the models and the observations,  
in that the whole dynamical 
range charachterizing the measured SFRs seems to roughly correspond to the values 
spanned by the models. 
The same is not true as far as the dust mass range is concerned, in particular for the COSMOS/GOODS sample:
starting from log[SFR/($M_{\odot}/yr$)] $\sim$ 1  and moving rightwards, many data points 
present dust masses underestimated by our models calculated with a standard IMF, 
in particular in the range $1\lesssim$log[SFR/($M_{\odot}/yr$)]$\lesssim 3$. 

The tracks of the PSPHs show an early steep increase of the $M_{dust}$-SFR relation, 
a maximum dust mass achievied still at early times and in correspondance of the largest SFR values, 
followed by a decrease of both quantities at later epochs. This trend is particularly strong for the largest mass system. 
This result is qualitatively in agreement with Hjorth et al. (2014), who found a 'maximal dust mass' 
which can be achieved in strongly star forming systems. 
In particular, our models allow us to appreciate the dependence of this maximal dust mass on the star 
formation history of PSPHs and spirals, and how this quantity basically increases 
as a function of the SFR and, in PSPHs, increases also at fixed SFR by assuming a THIMF. 
In fact, as visible from Fig~\ref{fig0}, the assumption of a THIMF has little effect on the star formation history of PSPHs. 
Since a Schmidt (1959) law is used for the SFHs of our models, this occurs as the gas mass budget is not dominated by the gas returned from stellar populations, which depends strongly on the assumed IMF,  
but by the amount of accreted pristine gas. 

Not surprisingly, also in this case the inclusion of a THIMF in PSPHs reduces considerably the tension 
between the most extreme data and models, 
as it has little effect on the star formation histories but a strong impact on the predicted dust masses.

\subsection{Dust mass vs gas mass}
\label{mdust_mgas}

The samples of star forming galaxies considered in the present work 
are characterised by a positive correlation between the dust mass and the gas mass. 
The values observed in most low-redshift discs are compatible with the tracks of the spiral models. 
Even if for a few ULIRGs the  $M_{dust}-M_{gas}$ relation is marginally consistent with the one found for the 
intermediate mass PSPH model, at variance with the $M_{dust}-$SFR plot, 
in the $M_{dust}-M_{gas}$ plot most of the ULIRGs with available gas masses lie between the lower mass and intermediate mass PSPH model, 
i.e. the PSPH models overestimate the gas masses of the ULIRGs sample. 
The HI mass in ULIRGs is not observed and, as discussed in previous works (Sanders \& Mirabel 1996; Santini et al. 2010),  
the presence of a large, undetected HI gas reservoir in ULIRGs seems implausible. 
Although to a lesser extent, the same might be true also for the $M_{dust}$-$M_{gas}$ relation found for most of the SMGs, as compared 
to the one predicted for the most massive PSPH model. 
These facts indicate that the most massive PSPH models tend to overestimate the molecular gas masses observed in ULIRGs and SMGs. 

The PSPH models show that 
after a rapid increase of the dust buildup which, as seen in Sect. ~\ref{mdust_sfr}, 
occurs at nearly constant SFR, the dust masses reach a maximum and then 
start decreasing in lockstep with the SFR, as the available gas reservoirs undergo  progressive consumption. 
During this late phase, the models move along a nearly 
diagonal line which starts from the maximum of the dust mass ever achieved during the whole 
evolutionary pattern,   
and evolves towards lower $M_{dust}$ and $M_{gas}$ values, i. e. 
towards the point which marks the end of star formation. 
This point is located on the lower left with respect to the $M_{dust}$ maximum. 

This phase in generally longer for PSPHs with larger stellar mass, owing to 
a larger star formation efficiency and a faster gas consumption timescale, 
which imply a steeper decrease of the SFR and dust mass vs time.

Model results calculated assuming a THIMF in PSPHs are 
presented also in the lower panel of Fig.~\ref{fig1}, which allow us to explain  
the behaviour of the most-extreme points of the datasets, i. e. the few ULIRGs and SMGs 
characterised by particularly large dust masses at particularly low gas mass values. 
This is not the only way to reduce the discrepancy between data and models 
in this diagram and in the others previously described in Sect.~\ref{mdust_mstars} and 
~\ref{mdust_sfr}. 
The effects of other processes will be discussed later in Sect.~\ref{disc}.

\begin{table*}
\begin{tabular}{lcccc}
\\\hline 
  Sample     &   log($M_{star}/M_{\odot}$)     &   log($M_{dust}/M_{star}$) &                       &                        \\    
             &                               &    Median value           &  16$^{th}$ percentile   &  84$^{th}$ percentile   \\   
\hline 
SINGS        &                               &                                                  \\   
\hline 
              &   10.60                      &  -3.22    &  -3.40  & 3.00                        \\
              &   11.13                      &  -4.07    &  -4.25  & 3.57                        \\  
\hline
KINGFISH      &                              &                                                    \\  
\hline 
              &    8.78                      &   -2.69   &   -3.70  &    -2.38                     \\
              &    10.37                     &   -3.34   &   -4.35  &    -2.74                     \\
\hline
COSMOS/GOODS  &                              &                                                     \\
\hline

              &      10.07                   &     -2.40 &   -2.83  &    -1.83                     \\
              &     10.47                    &     -2.52 &   -2.87  &    -2.10                     \\
              &     10.76                    &     -2.56 &   -2.91  &    -2.15                     \\
              &     11.03                    &     -2.73 &   -3.08  &    -2.32                     \\
              &     11.36                    &     -2.83 &   -3.16  &    -2.35                     \\
                                                                                                  
\hline 
ULIRGs        &                              &                                                      \\
\hline
             &    10.91                      &   -2.65  &    -2.98  &   -2.21                       \\
             &    11.66                      &   -3.12  &    -3.70  &   -2.15                       \\
\hline 
SMG          &                                &                                                      \\
\hline
             &     10.82                      &   -1.85  &     -2.12 &     -1.28                     \\
             &     11.21                      &   -2.40  &     -2.97 &     -2.26                     \\
\hline
Calura et al. (2014) &                       &                                                      \\
\hline
              &    10.56                      &   -1.95  &    -2.65   &   -1.45                      \\
\hline
\end{tabular}
\caption{Median dust-to-stellar mass ratio as a function of the stellar mass calculated for the samples described in Sect.~\ref{sec_obs}. First column: name of the sample. 
Second column: median value of the stellar mass distribution in each bin. Third column: median DTS ratio in the bin. Fourth and fifth columns: 16$^{th}$ and 84$^{th}$ percentiles of the DTS ratio distribution in the bin, respectively.}
\label{tab2}
\end{table*}

\begin{table*}
\begin{tabular}{lcccc}
\\\hline 
  Sample     &   $<z>$     &   log($M_{dust}/M_{star}$)       &                        &                      \\
             &             &    Median value                &  16$^{th}$ percentile   &  84$^{th}$ percentile   \\   
\hline        
SINGS        &            &                                &                   &                                 \\
\hline        
             &   0.00     &        -3.40                   &        -4.25      &   -3.08                     \\
\hline                                                         
KINGFISH     &           &                                 &                   &                                 \\         
\hline                              
             &   0.00    &         -3.03                   &        -3.81      &   -2.54                     \\
\hline                                                              
COSMOS/GOODS&            &                                &                   &                               \\                                  
\hline                                                         
             &   0.18    &         -2.81                 &         -3.17      &    -2.44                      \\
             &   0.42    &         -2.66                 &         -3.00      &    -2.32                      \\
             &   0.74    &         -2.57                 &         -2.93      &    -2.13                      \\
             &   1.01    &         -2.53                 &         -2.92      &    -2.04                      \\
             &   1.64    &         -2.49                 &         -3.06      &    -1.90                      \\
\hline                                                         
ULIRGs      &            &                                &                   &                                \\                                  
\hline                                                         
            &     0.11   &         -2.83                 &         -3.33      &   -2.34                        \\
\hline                                                         
SMG         &             &                              &                    &                                   \\                                  
\hline                                                         
            &     1.22   &         -2.30                 &         -2.97      &   -1.28                         \\
            &     2.52   &         -2.26                 &         -2.90      &  -1.94                          \\
\hline                                                         
Calura et al.  (2014)    &                               &                    &                               \\                                  
\hline                                                         
             &    5.95   &         -1.81                 &         -2.41      &    -1.44                       \\
\hline 
\end{tabular}
\caption{Median dust-to-stellar mass ratio as a function of redshift calculated for the samples described in Sect.~\ref{sec_obs}. First column: name of the sample. 
Second column: median value of the redshift distribution in each bin. Third column: median DTS ratio in the bin. 
Fourth and fifth columns: 16$^{th}$ and 84$^{th}$ percentiles of the DTS ratio distribution in the bin, respectively.}
\label{tab3}
\end{table*}

\subsection{Dust-to-stellar mass ratio vs stellar mass and redshift}
\label{dts}
A useful quantity to assess the efficiency of the dust 
production in galaxies is the dust-to-stellar mass (DTS) ratio. 
This quantity accounts for 
the amount of dust per unit stellar mass which survives the various 
destruction processes in galaxies and is effectively observable. 

\begin{figure*}
\includegraphics[width=180mm,height=180mm,angle=0]{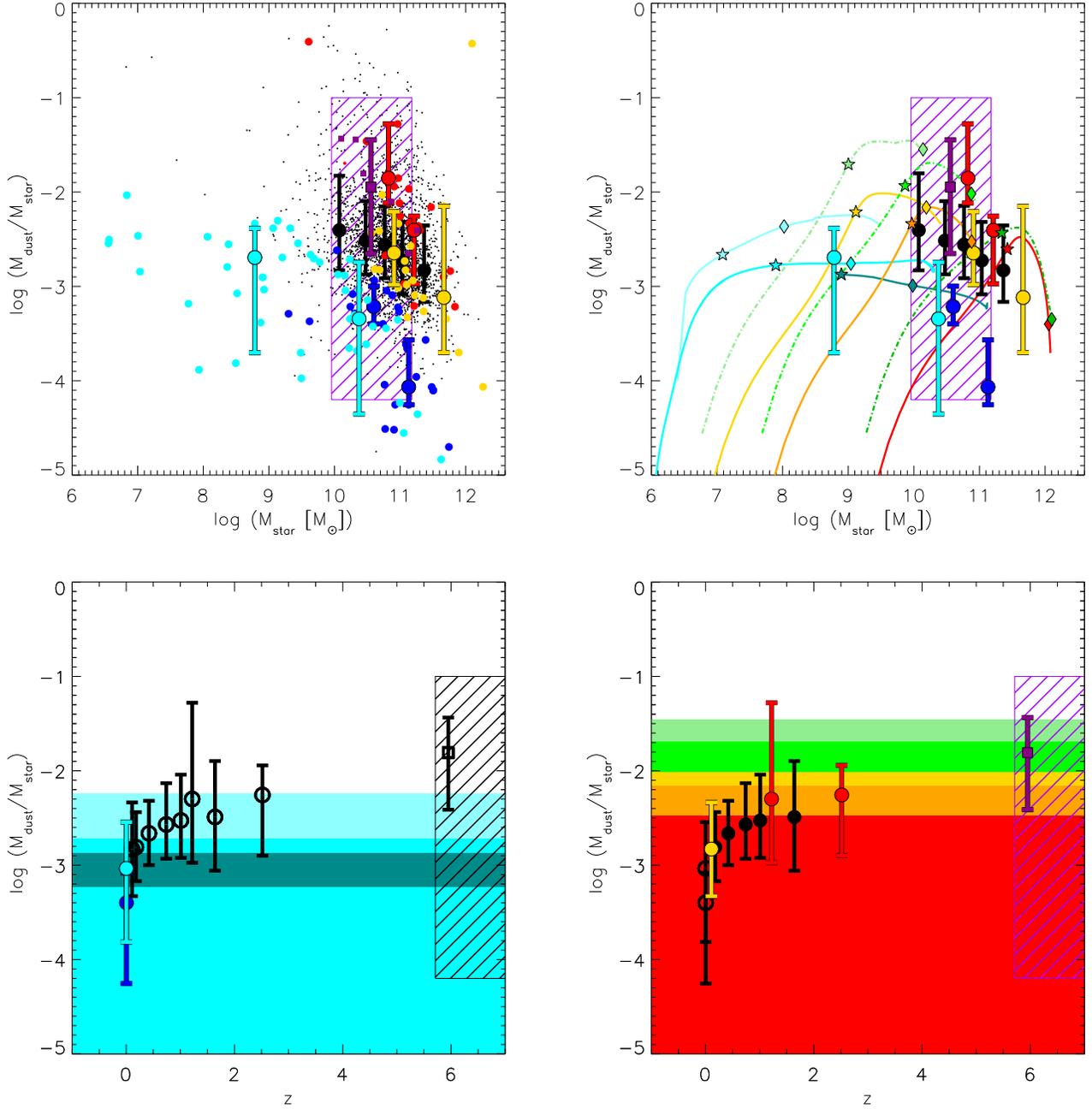}
\caption{Top panels: dust-to-stellar mass ratio as a function of the stellar mass. 
The large solid cyan, blue, yellow, red and black circles plotted with the error bars 
represent the median DTS values calculated in bins of stellar mass for the KINGFISH, SINGS, ULIRG, SMG, and COSMOS/GOODS samples, respectively.
The small purple squares are the DTS values measured in 8 QSO hosts from the sample of systems at $z\sim 6$ of 
Calura et al. (2014), 
i.e. in the systems for which it was possible to derive an estimate of the stellar mass. Note that in these systems, 
the stellar masses were computed from measures of the dynamical masses and neglecting the presence of dark matter and that if the latter was present, 
they should be regarded as upper limits. 
The large purple square with the error bar represents the median DTS value of the sample of Calura et al. (2014), 
as computed for these 8 systems. The purple hatched area accounts for the non-detections of the  $z\sim 6$ sample (see Sect.~\ref{dts} for details on its calculation). 
All the other curves and symbols are as in Fig.~\ref{fig1}. All the median DTS values as a function of the stellar mass are reported 
in Tab.~\ref{tab2}.
Bottom panels: dust-to-stellar mass ratio as a function of redshift. 
In the bottom-left panel, the solid cyan and blue circles are 
the median DTS values calculated in various redshift bins for the local KINGFISH and SINGS samples, respectively, 
whereas the open symbols represent the median values calculated for higher-redshift samples (see description of the bottom-right panel). 
The light-cyan, cyan and dark cyan regions in the left-bottom panels 
enclose the range of DTS values predicted for low-mass, intermediate mass and larger mass spiral galaxies, respectively. 
In the bottom-right panel, all the symbols with error bars are as in the top panels. 
The purple hatched area accounts for the non-detection of the sample of systems at $z\sim 6$ of Calura et al. (2014); 
its width was calculated from the difference between the two extreme redshift values of the objects belonging to that sample.}. 
The yellow, orange and red regions in the bottom-right panel enclose the DTS values of the PSPH models of lower mass, intermediate mass 
and larger mass, respectively, calculated assuming a Salpeter (1955) IMF. The light-green and green regions in the right-bottom panel 
enclose the DTS values computed for the lowest mass and intermediate mass PSPHs, respectively, 
computed assuming a THIMF. 
All the average DTS values as a function of redshift are reported in Tab.~\ref{tab3}. 
\label{fig3}
\end{figure*}

In the top panels of Fig.~\ref{fig3}, we show the DTS as a function of 
the stellar mass as observed in galaxies at various redshifts, and 
corresponding model tracks of our chemical evolution models for galaxies of
different morphological types. 

The observed relations are plotted along with the median values, i. e. the 50th percentiles of 
the distributions computed in bins of stellar mass, represented by the large solid circles. 
The bin width has been chosen in order to have comparable numbers of objects in each bin. 
In each stellar mass bin, the error bar represents the dispersion associated to the median value, 
and connects the 16th to the 84th percentiles of the observed distribution. 
The median values and their associated dispersions computed for each observational dataset are 
also presented in Table~\ref{tab2}. 

In general, the observational data sets indicate an anti-correlation between the DTS and 
the stellar mass. 
This is in agreement with previous observational studies of the DTS as a function of 
the stellar mass in nearby galaxies in different environments (Cortese et al. 2012) and 
in GRB hosts (Hunt et al. 2014). 
This result implies that in low mass galaxies, 
the specific production of dust is particularly efficient and that the 
balance between dust production 
and destruction is somehow dependent on the mass of the galaxy (see Sect.~\ref{disc}). 
We have verified that this trend is not due to any redshift effect. 
By calculating the DTS-mass relation in various redshift bins, 
a similar anti-correlation was found in each bin, characterised by 
a systematic shift towards higher DTS values with increasing redshift at all masses. 

The shapes of the theoretical evolutionary tracks (top-right panel of fig.~\ref{fig3}) outline the strong dependence of 
this quantity on the star formation history, which in turn depends on other quantities,  
such as the baryonic infall rate, the dust-destruction rate and the rate of astration. 

In general, in systems characterised by a nearly constant and prolonged star formation 
activity, the DTS shows a rather flat behaviour with respect to stellar mass. 
This is visible from the tracks calculated for spiral galaxies. 
For these systems, the three curves run along nearly horizontal tracks at 
different DTS levels.  

All the curves calculated for PSPHs show a rather different behaviour. 
During the earliest phases they 
grow steeply up to a maximum DTS level, 
after which they start decreasing until the end of SF. 

The KINGFISH and SINGS data are in good agreement with the spiral model curves, whereas 
the ULIRGs data are well-reproduced by the intermediate- and high- mass PSPH model. 
The median values calculated for the COSMOS/GOODS data are in general consistent with the 
PSPH tracks. 
Moreover, the tracks computed with a standard IMF underestimate some of the observed DTS values, i.e.  
the lowest-mass SMG value and the high-redshift value from the sample of Calura et al. (2014). 
This discrepancy between models and observed values vanishes once a THIMF is assumed. 
Finally, it is worth to note that an anti-correlation between he DTS and 
the stellar mass is also found in a recent study of McKinnon et al. (2106), based on cosmological simulations including dust production. 

The redshift evolution of the DTS ratio is shown in the lower panels of Fig.~\ref{fig3}, 
compared with the values predicted by means of our models for spiral discs (left-bottom panel) and PSPHs (right-bottom panel). 
The data 
are presented in Table~\ref{tab3} and 
represent the median values of the observed datasets. 
The median DTS values were computed in different redshift bins and 
are plotted along with their dispersion, which has been calculated 
from their 16th and 84th percentiles as done for the data in the upper panels and in Table~\ref{tab2}. 

The data suggest an increasing trend of the DTS as a function of redshift as visible in the range $0\le z \lesssim 2.5$. 
At larger redshift, there could be a wide spread in the dust content of star-forming galaxies (e.g. De Cia et al. 2013), 
which might also result in a large spread in the DTS ratio. 

At $z\sim6$, the detections of the sample of Calura et al. (2014) indicate DTS values larger than those at $z\sim 2$. 
However, these systems are likely to represent the most luminous and the dustiest objects of this sample. 
The hatched area plotted in Fig.~\ref{fig3} accounts for the several non-detections of the $z\sim 6$ sample 
and for uncertainties in the stellar mass estimates. 
It is worth stressing that, strictly speaking, the stellar masses should be regarded as upper limits 
and consequently the DTS ratios represent lower limits. 
 
An increasing trend of the DTS-redshift relation was also found by Dunne et al. (2011), although in a much narrower 
redshift range ($0\le z \le 0.3$).  
We have verified that this trend is not due to the different range of stellar masses covered at various redshifts  
by dividing the largest sample considered in this paper, i.e. 
the COSMOS/GOODS  sample, into various mass bins, and computing in each mass bin the median DTS as a function of redshift, 
dividing the sub-sample into redshift bins. 
In each mass bin,  we still find a DTS increasing with redshift, 
with a hint of a steeper increasing with redshift for lower mass galaxies. 
The observed DTS increases roughly by 0.5-0.7 dex from $z\sim 0$ to $z\sim 1$, 
to reach a plateau in the interval $1\le z \le 2$, where the galaxies of the COSMOS/GOODS sample lie. 
The SMG sample extends up to larger 
redshifts, and present median DTS values slighly larger (by $\sim$ 0.2 dex) than the ones obtained for the COSMOS/GOODS sample 
at comparable redshift.\\
The 8 systems at $z\sim 6$ present very large DTS values, by 0.7-0.8 dex larger than the plateau of the COSMOS/GOODS galaxies between $z\sim 1$ and $z\sim 2$. 
The purple hatched area calculated at $z\gtrsim6$ takes into account the non-detections of the sample of Calura et al. (2014), 
and is compatible with a constant or decreasing DTS value at high redshift. 
The redshift range spanned by the non-detections could embrace an early epoch in which a decreasing trend of the global DTS 
is more likely than an increasing one, as expected 
from the theoretical time evolution of the DTS in both PSPHs and spirals shown in Fig.~\ref{fig0}. 
In particular, our PSPH models suggest a steep rise of the galactic dust content, which takes place on timescales of the order $\sim 0.1$ Gyr, and 
with a maximum dust mass reached earlier in more massive systems. 
Such a steep rise of the dust content in the early Universe has already been the topic of a few recent publications.  
In fact, while the 8 objects at $z\sim 6$ show a large presence of dust in extremely luminous (SFR$>>10$ M$_{\odot}$/yr) starbursts hosting QSOs, and 
while at $z>6$ also dusty 'normal' star-forming galaxies (i.e. forming stars from a few to a few tens of solar masses per year) have been found 
(see Mancini et al. 2015), a few notable cases of dust-free galaxies have been discovered at $z>7$ (Ouchi et al. 2013; Inoue et al. 2016). 
Their existence may be explained by means of a very rapid evolution of the dust content of the Universe  at $z>6$, and with a sudden appearance of large 
amounts of dust as soon as adequate reservoirs of refractory elements become available (Mattsson 2016).\\

\section{Discussion}
\label{disc}
In general, the models provide an acceptable description of the observed scaling 
relations for most of the KINGFISH and SINGS galaxies, 
well described by the spiral galaxy models, and 
for the bulk of the ULIRGs, for the SMGs and 
COSMOS/GOODS sample, well described by the PSPH models. 

The dust masses derived for many COSMOS/GOODS galaxies are larger than 
the ones characterising the model galaxies. 
It is worth stressing that a few systematic effects could be present in the observational 
estimate of the dust masses, possibly leading to overestimations.

\subsection{What process drives the DTS-mass relation?} 

Little is known about dust production in particularly intense star formation environments;
the range of model parameters involved is wide (dust condensation efficiency in AGB stars and supernovae, 
dust growth and destruction efficiency, which could all depend on metallicity and/or star formation rate). 
In order to build a realistic model of dust production in such an environment, 
one should properly take into account 
the dependence of growth and destruction on a few key physical properties of the the interstellar gas, such as density and temperature, 
as well as a few non-thermal properties, such as 
microscopic turbulence or magnetic fields. Currently, these facts render a 
credible "ab initio"  modeling of interstellar dust far from feasible. 

The chemical evolution models used in this work 
are useful to understand the global trends of the studied relations, as well as the main macroscopic 
processes these  are driven by. 
We have seen that both the star formation history and the stellar mass play important roles in defining 
the DTS-M$_{star}$ relation. 
The observations indicate that lower mass systems, generally characterised by a less intense 
star formation activity, show larger DTS values than 
larger mass, more intensely star forming objects, broadly in agreement with model predictions. 

A decreasing trend of the DTS with the stellar mass reflects the time evolution of the theoretical DTS in PSPHs  
(see Fig.~\ref{fig0}) in the late starburst phase, i.e. the phase following the peak of the star formation epoch.
During this stage the star formation rate starts decreasing, as well as the dust production rate. 
As the stellar mass still continues to increase, the final result is a DTS ratio 
moving along a diagonal line towards the lower-right corner of the diagram, 
i.e. towards the point which marks the end of star formation. 
Regardless of the stellar mass, 
this late starburst phase, characterised 
by the decrease of the DTS ratio, is considerably more extended than the early 
phase during which the DTS ratio increases with time (see Fig.~\ref{fig0}). 
This implies that the DTS-M$_{star}$ relation 
is more likely populated by starbursts caught during this late phase. 

\begin{figure*}
\includegraphics[width=100mm,height=80mm,angle=0]{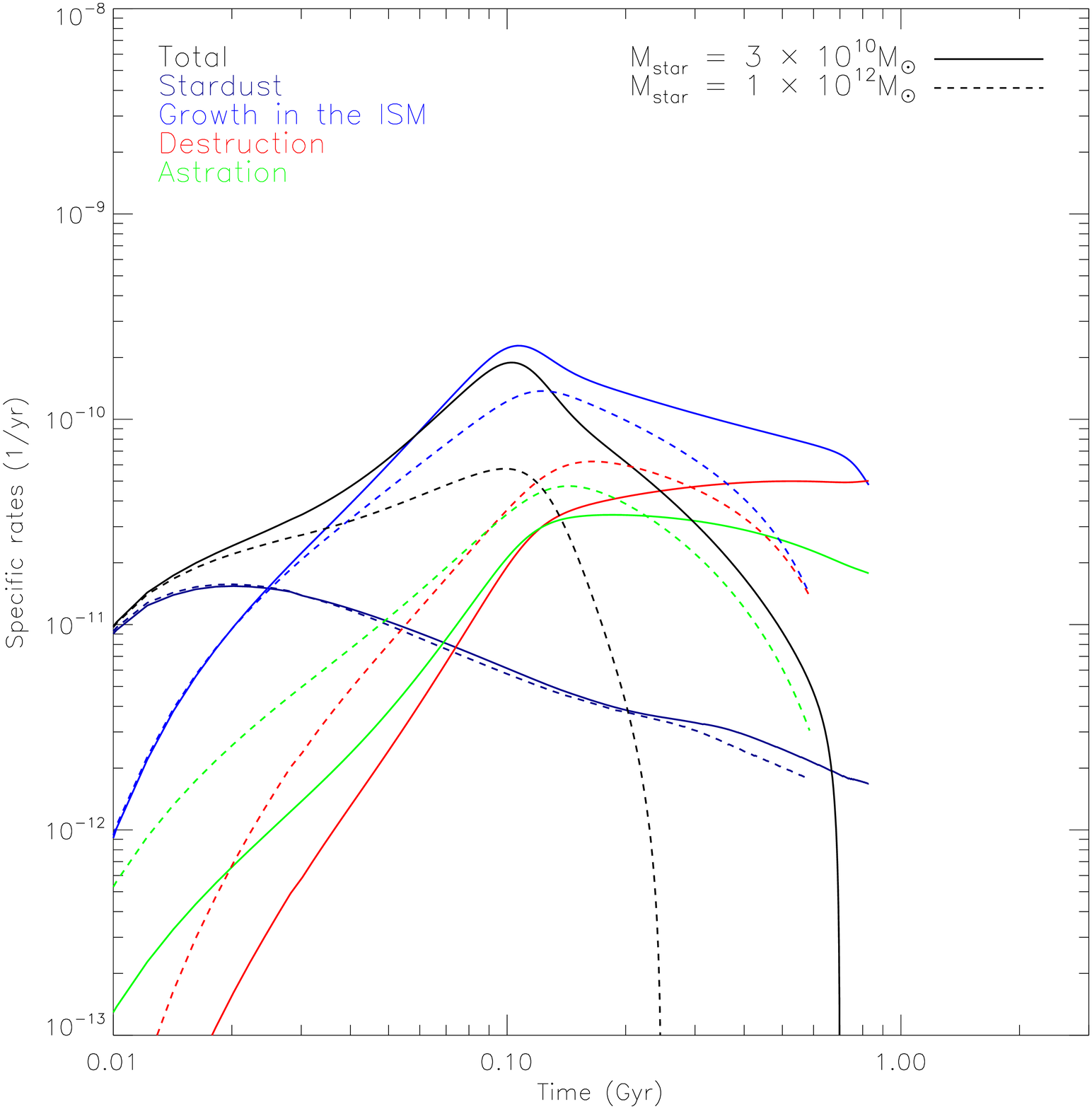}
\caption{Theoretical specific dust production and destruction rates as a function of time in proto-spheroids. 
Results for two models of baryonic mass $3 \times 10^{10}$ \msun\, and  $10^{12}$ \msun\, are shown (solid and dashed lines, respectively). In both cases, a standard Salpeter (1955) IMF is assumed. 
The green and red lines represent the destruction rates as due to astration and to shocks in supernovae, respectively. 
The blue, dark blue and black lines represent the interstellar dust growth rates, the stellar dust production rates and the total rates, 
respectively.}
\label{fig4}
\end{figure*}

Provided that the observed systems are all caught at comparable times after the beginning of the starburst, 
a DTS decreasing with stellar mass can also be explained by considering the theoretical 
DTS-time relation shown in Fig.~\ref{fig0} for PSPHs and that, at a fixed time, the DTS decreases with mass.\\
Moreover, the DTS is larger in low-mass systems because of the net specific rates of dust production 
which, similarly to the specific SFR (e.g., Hunt et al. 2014), tends to 
be larger in lower mass galaxies. 
This is summarised in Fig.~\ref{fig4}, where the specific 
dust production and destruction rates are 
shown as a function of time for PSPHs of M$_{star}=3 \times 10^{10}$ \msun\, and M$_{star}= 10^{12}$ \msun. 

Although the specific stellar dust production rates are comparable (dark blue solid and dashed lines), 
the lower mass model shows a lower specific destruction rate and a larger growth rate than the 
larger mass model. 
As shown in various previous studies (Dwek 1998; Calura et al. 2008; Pipino et al. 2011; Calura et al. 2014), 
the balance between destruction and growth is crucial to regulate the dust mass budget; 
in this case, we have shown the fundamental importance of these processes also in driving the DTS-M$_{star}$ relation.

\subsection{Why a top-Heavy IMF is needed}

In models, larger dust masses can be achieved by means of a stellar IMF skewed towards massive stars, i.e. 
a top-heavy IMF, as this implies a more efficient stellar dust production 
(Gall et al. 2011b; Calura et al. 2014). 

At the same time, even if the higher SN rates achieved with a THIMF imply larger destruction rates, 
also the accretion rate increases because of larger dust masses and  
of a lower dust-to-metal ratios (Calura et al. 2014). 
Moreover, the lower stellar masses achieved with a THIMF lead to 
dust masses produced per unit stellar mass larger than those obtained with a standard IMF. 

Although no direct, ultimate proof for a different IMF in the early 
Universe or in starbursts can be found (Larson 1998; Elmegreen 2005; Narayan \& Dav\'e 2012), 
a top-heavy IMF offers a possible explanation for the large dust masses observed in high-$z$ starbursts 
(Gall et al. 2011a,b; Dwek et al. 2011; Calura et al. 2014). 

The study by Marks et al. (2012) suggests an IMF becoming more top-heavy at low metallicities 
and at high interstellar gas densities. 
Moreover, in principle, in 
extremely strong starbursts with SFR $>1000$ \msun/yr, the formation of very massive and dense 
star clusters should be favoured, naturally leading to IMF skewed towards high mass stars (Weidner et al. 2011).  

Possible ways to explore the initial mass function in starbursts include multiwavelength observations 
aimed at probing the production rate of ionizing photons. 
Recent results based on studies of unresolved stellar clusters in nearby galaxies show no deviation 
from a universal, standard IMF (Andrews et al. 2013). 
Future instruments such as the James Webb Space Telescope (JWST) will allow to extend such studies to higher 
redshifts, and this will be fundamental to gain more direct information 
on the IMF in distant, intense starbursts. 

Finally, the study of interstellar metallicity offers 
another extremely efficient way to test 
the IMF in star-forming galaxies. 
In fact, as visible in Fig.~\ref{fig0},  
the interstellar metallicity achieved in PSPHs at the end of the starburst and calculated assuming a THIMF 
can be  more than
0.5 dex higher than that computed with a Salpeter IMF. 

The most dust-rich and strongly star-forming systems are expected to contain very large reservoir of metals. 
In such dust-enshrouded objects, metallicity diagnostics based on rest-frame optical emission lines are unreliable,  
as the optical spectral indices are likely unreliable (Santini et al. 2010). 
Currently, the Atacama Large Millimeter/submillimeter Array (ALMA) offers the possibility 
to overcome this problem. In fact, a recent ALMA study based on the flux ratio of the far-infrared fine-structure emission lines [NII]205$\mu$m
and [CII]158$\mu$m as metallicity diagnostics has revealed 
high levels of chemical enrichment even in a starbust at $z=4.76$ (Nagao et al. 2012).

\section{Conclusions}
In this paper, we have considered a few dust-related 
scaling relations in galaxies at both low and high redshift, 
in order to gain some basic information on how star formation 
has progressed in galaxies of various 
morphological types as a function of cosmic history.  
The scaling relations studied here include dust mass, gas mass, stellar mass and SFR. 

We have assembled various data sets at 
various redshifts, and which include galaxies characterised by a large variety of 
star formation histories and spanning a large dynamical range in both stellar mass and SFR. 
These data sets include local spiral discs such those from the SINGS and KINGFISH projects, 
as well as data from the COSMOS/GOODS surveys, which span a wide 
redshift range and include many intense starbursts, as well as local ULIRGs and of SMGs at $z > 1$. \\
The data sets have been interpreted by means of chemical evolution models 
of galaxies of different morphological types, useful in particular 
to set a few constraints on the star formation history of the observed systems 
and to understand which physical processes drive the relations 
considered in this paper. 

We have focused in particular on the amount of dust per unit stellar 
mass observable in local and distant discs and starbursts, 
which represents a measure of the dust production efficiency or, rather, 
of the survival capability of dust grains in quiescent galaxies and in starbursts. 
We study how this quantity evolves as a function of the stellar mass and 
redshift and how it can be theoretically accounted for. 

Our main results can be summarised as follows. \\

{\it(i)} 
The dust content of local disc galaxies and ULIRGs is broadly accounted for by 
our chemical evolution models for spirals and PSPHs, respectively. 
On the other hand, in several cases the large dust masses observed in some high-redshift samples 
are difficult to reproduce by means of our models with standard assumptions on basic parameters such as, e.g., 
the stellar IMF. 
This fact outlines a notorious dust budget problem already addressed in previous studies 
(Dwek et al. 2007;  Gall et al. 2011a; Mattsson 2011; Rowlands et al. 2014; Hjorth et al. 2014). 
A Top-Heavy IMF in PSPHs allows us to alleviate  the tension 
between data and models. However, in a few particularly dust-rich systems 
other processes could play an important role, 
such as an enhanced dust growth efficiency (Valiante et al. 2011; Asano et al. 2013; Nozawa et al. 2015). \\

{\it(ii)} The study of the relation between 
dust mass and SFR in high redshift samples allows one to extend 
this diagram to SFR values larger than those measured in local starbursts. 
In high redshift samples, the $M_{dust}$-SFR relation 
appears flatter than the one derived in local 
star-forming galaxies by Da Cunha et al. (2010). 
In fact, the maximum dust masses observed in high-redshift samples are of the same order 
as those observed locally, but the SFRs of high-redshift systems are generally larger.  
The model results indicate that the two morphological types tend to occupy rather distinct regions of this diagram: 
spiral discs occupy the lowest dust mass, lowest SFR part, whereas 
PSPHs occupy the opposite part of the plot, characterised by 
the highest-SFR and largest $M_{dust}$ values. 
Overall, the samples considered in this work which bear information on the morphology of the galaxies 
tend to reflect this trend. 
In agreement with Hjorth et al. (2014), our models indicate that there is a maximum 
dust mass achievable in systems with log[SFR/($M_{\odot}/yr$)]$>3$. 
Our models show the dependence of this ``maximal dust mass'' on the star 
formation history of PSPHs and spirals. This quantity tends to increase 
as a function of the SFR and, in PSPHs, increases also at fixed SFR if one assumes a THIMF. \\

{\it(iii)} The $M_{dust}$-SFR and the $M_{dust}$-$M_{gas}$ correlations 
are populated by starburst galaxies caught in their latest star-forming phases. 
The PSPH models show that 
after a rapid increase of the dust buildup, 
occurring at nearly constant SFR, the dust masses reach a maximum and then start decreasing as the SFR. 
During this late star-forming phase, each model moves along a nearly 
diagonal line which starts from the maximum of the dust mass ever achieved during the whole 
evolution,    
and evolves towards lower $M_{dust}$ and $M_{gas}$ values, i. e. 
towards the point which marks the end of star formation.\\

{\it(iv)} The observational data sets indicate an anti-correlation between the DTS and 
the stellar mass, implying  that the production of dust per unit stellar mass is particularly 
efficient in low mass galaxies. 
The theoretical models demonstrate the strong dependence of 
this quantity on the star formation history.
In systems characterised by a nearly constant and prolonged star formation 
activity, the DTS shows a rather flat behaviour with respect to stellar mass. 
On the other hand, in the earliest phases of PSPHs the DTS grows steeply up to a maximum DTS level, 
after which it experiences a decrease which extends up to the end of SF.
Once again, our models computed with a standard IMF underestimate the DTS observed in distant 
starburst galaxies; the adoption of a THIMF helps reducing significantly the tension between data and models.\\

{\it(v)} The decreasing trend of the DTS with stellar mass reflects the late starburst 
phase of PSPH models, i.e. the phase following the peak of the star formation epoch, characterised 
by a constant decrease of both the SFR and of the net rate of dust production. 

In the mean time, the continuous increase of the stellar mass causes the 
DTS to proceed along a diagonal line towards the end of star formation. \\
The DTS ratio is larger in low-mass PSPHs because these are characterised 
both by a lower specific destruction rate and by a larger specific growth rate than the 
larger mass models. \\

{\it(vi)} The observed DTS ratio shows an increasing trend with redshift 
in the range $0\le z \lesssim 2.5$. 
It increases roughly by 0.5-0.7 dex from $z\sim 0$ to $z\sim 1$, 
to reach a plateau in the interval $1 \le z \le 2$. 
The values  computed for the detections of the sample at $z\sim 6$ are within the error bars to those of SMGs. 
However, at $z\gtrsim 6$ the spread in the DTS is expected to be wide. 
Our models predict that 
at some point at larger redshift, the global DTS is expected to decrease with increasing redshift. 
The growth of dust in the early Universe could be extremely rapid 
(Mattsson 2016), as supported also by the few notable cases of dust-free galaxies found at $z>6.5$ (Ouchi et al. 2013; Inoue et al. 2016). \\
The average DTS computed from the data of local discs and ULIRGs are comparable with theoretical estimates 
for spiral galaxy models and PSPHs, respectively. 
Some of the observational estimates at $z>1$ are reproduced by our PSPH models which adopt a THIMF. 
This could be another indication that a THIMF is required to explain the large dust masses in high-redshift data. 
However, such an assumption leads to metallicities larger than the ones computed with a standard 
IMF by more than 0.5 dex. Future infrared studies based on the analysis of 
flux ratio of fine-structure emission lines  as metallicity indicators (Nagao et al. 2012) 
will shed light on the metal content of high$-z$ starbursts. 

In general, a better understanding of the grain size distribution and temperatures, as well as their composition will be 
fundamental in order to improve the accuracy in observatinal estimates of the dust mass. 
In particular, a valuable quantity useful to constrain the size distribution is the extinction curve, 
which depends also on the dust chemical composition. In principle, the study of the extinction curves in galaxies whose chemical 
abundance pattern can be derived by means of different observables, such as in the Milky Way or in the Small Magellanic Cloud, can be very useful 
to constrain the size distribution (e. g. Schurer et al. 2009). 

On the theoretical side, 
it will be important to match the predictions from chemical evolution models including dust production to spectro-photometric 
codes such as GRASIL (Silva et al. 1998); previous efforts in this direction are those of Schurer et al. (2009) and Pozzi et al. (2015). 
In this regard, significantly new perspectives are also offered by the GRASIL-3D tool (Dom\'inguez-Tenreiro et al. 2014), 
designed to match a detailed spectro-photometric modelling of dust, including radiative transfer effects, 
to the outputs of hydrodynamical galaxy formation codes. 

Another future step will be to investigate the evolution of the dust content in galaxies 
by means of models computed within a cosmological framework. 
An attempt to model dust production in a fully cosmological context was made by 
McKinnon et al. (2016), which provides detailed three-dimensional theoretical distributions of the dust in simulated galaxies. 
However, the best tools to provide a cosmology-based, 
theoretical description of the global properties of galaxies are semi-analythic models (SAMs) of galactic evolution. 
Ideally, in order to extend their predictive power to dust-related properties, next-generation SAMs will have to include a detailed treatment of the chemical evolution of 
refractory elements (e. g. Calura \& Menci 2009; 2011), of dust creation and destruction (e.g. Valiante et al. 2014) as well as 
a spectro-photometric treatment of dust grains. 

\section*{Acknowledgments}
We thank the anonymous referee for valuable suggestions that 
improved the quality of our work. 
Loretta Dunne is acknowledged for interesting discussions. 
PS acknowledges financial support from the European Union Seventh Framework 
Programme ASTRODEEP (FP7/2007- 2013), grant agreement n 312725.

\label{lastpage}

\end{document}